\documentclass[conference]{IEEEtran}
\usepackage{cite}
\usepackage{amsmath,amssymb,amsfonts}
\usepackage{algorithmic}
\usepackage{graphicx}
\usepackage{textcomp}
\usepackage{xcolor}
\usepackage{multirow}

\def\BibTeX{{\rm B\kern-.05em{\sc i\kern-.025em b}\kern-.08em
    T\kern-.1667em\lower.7ex\hbox{E}\kern-.125emX}}
\begin{document}

\title{Incorporating Classifier-Free Guidance in Diffusion Model-Based Recommendation\\
}

\author{\IEEEauthorblockN{Noah Buchanan, Susan Gauch, Quan Mai}
\IEEEauthorblockA{\textit{Department of Electrical Engineering and Computer Science} \\
\textit{University of Arkansas}\\
Fayetteville, USA \\
\{njb004, sgauch, quanmai\}@uark.edu}
}

\maketitle

\begin{abstract}
This paper presents a diffusion-based recommender system that incorporates classifier-free guidance. Most current recommender systems provide  recommendations using conventional methods such as collaborative or content-based filtering. Diffusion is a new approach to generative AI that improves on previous generative AI approaches such as Variational Autoencoders (VAEs) and Generative Adversarial Networks (GANs). We incorporate diffusion in a recommender system that mirrors the sequence users take when browsing and rating items. Although a few current recommender systems incorporate diffusion, they do not incorporate classifier-free guidance, a new innovation in diffusion models as a whole. In this paper, we present a diffusion recommender system that augments 
 the underlying recommender system model for improved performance and also incorporates classifier-free guidance. Our findings show improvements over state-of-the-art recommender systems for most metrics for several recommendation tasks on a variety of datasets. 
 In particular, our approach demonstrates the potential to provide better recommendations when data is sparse.
\end{abstract}

\begin{IEEEkeywords}
recommender system, generative AI, diffusion, classifier-free guidance
\end{IEEEkeywords}

\section{Introduction}
The introduction of generative models has marked a turning point in the capabilities of AI and it gives us a glimpse into the distances we can go using such tools. Autoencoders, a foundational generational technology, learn a compressed format of information to pass into neural models. However, the latent space created by the autoencoder's compression may contain parts that do not correspond to any data point in the original input, limiting their generative capabilities. Because the latent space is not regularized, autoencoder models only have generative capabilities that correspond to generating data we have seen before. During inference, when the training data is not available, we have no way of knowing whether or not the output generated is garbage.  To address this, Variational Autoencoders (VAEs) \cite{b28}
and Generative Adversarial Networks (GANs)\cite{b31} were developed, 
both of which have been shown to be capable of a wide range of  of generation capabilities.\cite{b31}
\cite{b28}
\cite{b30}.
Text-to-image synthesis is one such capability and, not coincidentally, a very popular area of research. The idea of a pipeline capable of transforming our human language into another form of media that has none of the ease of creation that language does is an extremely appealing concept; as a result, there has been an explosion of research in this area in recent years. GANs and VAEs were the main generative models used for a broad scope of tasks for some time. Despite their success, GANs and VAEs remain only partially competitive with recent state-of-the-art techniques - diffusion. 

Diffusion-based techniques outperform other recent techniques in the visual fidelity of images and their understanding of textual prompts.  Additionally, they excel in the range of generated images and their ability to create variations of the same image while still accurately following the guidelines provided by the text prompt. Diffusion models (DMs) are based on the idea of diffusion in the natural sciences and are heavily inspired by non-equilibrium thermodynamics. The process of diffusion in natural science is the movement of solutes and molecules from a high concentration to a lower concentration. As the name suggests, DMs attempt to mimic this process using Gaussian noise instead as our measure of concentration, a fully noised image being the parallel of a perfectly diffused liquid. 
The underlying concept is that, in actual diffusion, regardless of where or how dye is introduced into a container of water, the dye will eventually diffuse evenly throughout the container, producing a consistent result each time. DMs replicate this process, but instead of producing a uniform distribution, they recreate a distribution of the original data from random Gaussian noise.
Leveraging this principle, diffusion-based techniques have achieved unparalleled results in text-to-image synthesis and general image generation. While text-to-image is likely the most popular use case, they are capable of other tasks, such as text-to-text (sequence-to-sequence) or text-to-video. One such use case is recommendation. 

The current challenge with generative recommender models lies in the limited capablities of current approaches that primarily rely on GANs and VAEs. GANs are plagued by instability issues \cite{b29}, whereas VAEs face a trade-off between tractability and the quality of their representations \cite{b30}.
In contrast, Diffusion Models (DMs) do not encounter the aforementioned issues and therefore represent a promising solution for generative recommender systems (RSs). However, the current implementation of diffusion RSs is limited by their failure to incorporate recent innovations in DMs, i.e., classifier-free guidance. Essentially, classifier-free guidance, as implied by its name, enhances the generative capability of DMs without the need for an external classifier. Our contributions are threefold. \textit{First}, we develop an implementation that seamlessly integrates classifier-free guidance into diffusion-based RSs. \textit{Second}, we provide an in-depth analysis of the architecture of the employed models. \textit{Third}, we offer insights and solutions to the challenges posed by few-shot and zero-shot scenarios, which frequently hamper the performance of RSs.

Section \ref{sec:background} provides a summary of related work, situating our contributions within the existing literature. In Section \ref{sec:methodology}, we detail the process of integrating Classifier-Free Guidance into the implementation of a diffusion RS, originally described in our inspiration paper, and explain the inner workings of our enhanced RS. Section \ref{sec:results} presents the results of our modifications and compares them to the outcomes from unmodified implementations. Finally, Section \ref{sec:conclusion} offers conclusions and discusses potential avenues for future research to further advance the capabilities of diffusion-based RSs.

\section{Background and Related Work}\label{sec:background}
In this section, we discuss RSs and DMs and how these fields combine. We also give some background on a technique used in DMs that has seen significant success in recent literature, known as classifier-free guidance, which is the basis for the improvement demonstrated in this paper. 

\subsection{Recommender Systems}
RSs are a crucial part of how businesses operate today, with users providing information in return for recommendations on products, whether items from a store or shows/movies. The foundation for RSs was generally considered to be first introduced by Elaine Rich in 1979 through a system she called Grundy \cite{b1}. She wanted a way to recommend users' books by asking them specific questions, in turn classifying them into classes of preferences depending on their answers. Jussi Karlgren improved upon Elaine Rich’s system by introducing a measure of ``closeness'' to attempt to imitate the occurrences in bookcases where interesting documents often happen to be found adjacent to each other regardless of a relatively unordered bookcase \cite{b2}. This idea of closeness was adopted and used in a variety of models after its inception. The first information filtering system based on collaborative filtering through human evaluation was introduced by \cite{b3}, where they would make recommendations based on similarities between the interest profile of a user in question and other users; the concept was coined ``weaving an information tapestry.'' Inspired by the results of the study, researchers from MIT and the University of Minnesota (UMN) developed a new recommendation service called GroupLens, which features a user-to-user collaborative filtering model \cite{b20}. Following the publication of these initial findings, the GroupLens research lab was established at UMN and became a pioneering force in the field of recommender system studies. The GroupLens research lab subsequently launched the MovieLens project, which led to the creation of the first version of their recommender model \cite{b21}. 
As a result of this project and its success, several MovieLens datasets were released from then to now and became one of, if not the most, popular datasets for recommendation studies. 

AI-based methods have been very popular, with the explosion of AI research in the last 20 years being an obvious indicator of its soon-to-be application to the recommendation domain. The application in this field did see success as early as 2006 and 2009, inspired by the Netflix Prize, matrix factorization models that would learn a user embedding matrix and an item embedding matrix; using these embedding matrices, we can find similarities using the measure of closeness we talked about earlier, in this case typically it would be the dot product \cite{b22},\cite{b23}. Other techniques were created during this period, and linear models for this task gained some popularity. \cite{b24} presented a logistic regression model that achieved tangible improvements on new user-centric evaluation methods, such as errors in click-through rate estimation. \cite{b25} introduced a VAE for collaborative filtering with success on the MovieLens and Netflix Prize datasets. \cite{b26} showed that GANs are also capable of creating a recommender model. Models for video recommendation were also created following the introduction of deep learning for RSs \cite{b27}, showing that deep models have a surprising ability to recommend videos and the capability to perform both candidate generation and ranking. 


\subsection{Diffusion}
The first paper that broaches the topic of diffusion models coined “Deep Unsupervised Learning using Nonequilibrium Thermodynamics”, aimed to attack the tradeoff between tractability and flexibility that probabilistic models historically are afflicted by \cite{b4}. The next major breakthrough in DMs was achieved by the work in \cite{b5},  which marked the first significant adoption of this model for high-quality image generation. This study not only demonstrated that DMs are capable of producing high-quality images, competitive with state-of-the-art methods on the CIFAR-10 dataset, but also highlighted the efficacy of class-conditional models that use a classifier to guide generation towards higher fidelity on the same task. The influential findings of this paper catalyzed widespread adoption, leading to numerous subsequent improvements. DMs,  already known for the prowess in unconditional image generation, have shown superior performance in various other areas. Especifically, they have become the de-facto state-of-the-art in class-conditional generation, particularly with the incorporation of classifier guidance \cite{b7} \cite{b6}. Additional, DMs have begun to dominate the domain of super-resolution \cite{b8}. The advancements prompted by this major adoption extended beyond the models' use cases and capabilities to their performance optimization. Techniques such as latent diffusion models, introduced in the widely acclaimed Stable Diffusion paper \cite{b9}, have been developed to enhance these models. Latent diffusion models leverage the latent space concept from autoencoders: data is compressed into the latent space before the forward process, and both forward and reverse Markov transitions occur in this reduced dimensionality. The result is then upscaled back to the original size before output. This paper gained significant attention not only within the research community but also among non-researchers, as it provided open-access resources to code and model weights, enabling widespread use with proper guidance. Further improvements borrowed techniques from VAEs. For instance, \cite{b13} demonstrated that Vector Quantization in the latent space could enhance the efficiency of both training and sampling speeds. Another innovative approach by a team at Snapchat introduced step-distillation to decrease the number of steps required in the denoising process. This lead to the creation of the SnapFusion model, which boasts the capability of image generation within two seconds on mobile devices \cite{b14}. 

Large language models (LLMs) have represented a significant breakthrough in AI research, demonstrating exceptional performance across various applications, including chatbots, content generation, language translation, text summarization, question-answering systems, and personalized recommendations. This breakthrough has coincided with the rise in popularity of prompt-based conditional DMs, making it an opportune moment to merge LLMs' textual understanding capabilities with these advanced image generation models. Google explored this integration and discovered that large language models were surprisingly effective at encoding text for image synthesis. This led to the development of a model named Imagen \cite{b10}, which exemplified the potency of combining these techniques. Similarly, OpenAI's DALL-E 2 model employed large language models with diffusion techniques, using a CLIP model to generate images based on its enhanced language-image understanding \cite{b11}. Although the third iteration of DALL-E continued this approach, it uncovered some limitations; specifically, the imperfections in image captions were found to introduce flaws in the generated images. This highlighted the critical role of the language component in the effectiveness of these models \cite{b12}.

\subsection{Diffusion-Based Recommender Systems}
The success of DMs and the subsequent expansion of their use cases have led to increased interest in diffusion-based RSs. While GANs and VAEs are commonly utilized to model the generative process of user interactions, they share similar limitations with their traditional counterparts. With diffusion models showing promise in overcoming these issues within the domain of image generation, it is reasonable to hypothesize that they could address these shortcomings in generative recommender systems as well. The study by \cite{b15} provides compelling evidence in support of this hypothesis. This research introduced one of the first diffusion-based recommender models, arguing that the objectives of RSs align well with the methodological framework of DMs. Specifically, RSs aim to infer future interaction probabilities based on corrupted historical interactions, a process analogous to the noise-corruption mechanism inherent in DMs. Their approach did not involve any direct conditioning, instead, the amount of noise added to the data was limited so that the data was corrupted but not completely indecipherable.  

Although diffusion-based recommendation is a relatively new research area, it is rapidly gaining traction. Within the same year, several other diffusion recommender models emerged, employing varied approaches to achieve similar tasks. For instance, \cite{b16} introduced a conditional diffusion-based RS using a cross-attentive denoising decoder to remove noise. Unlike previous approaches that relied on guidance via partially corrupted data, this method employed direct conditioning. This inspired other papers, such as \cite{b18}, which developed sequential recommendation methods using DMs. Additionally, \cite{b17} explored the application of DMs for reranking purposes in RSs, achieving notable results.

\subsection{Classifier-Free Guidance}

\textbf{Classifier guidance} \cite{b6} modifies the diffusion score, specifically the gradient of the log probability density function, which is more tractable to learn than directly modeling the data distribution. This approach employs a classifier to guide the generation process by increasing the probability of data that the classifier assigns a high likelihood to the correct label. As demonstrated by \cite{b19}, data that is well-classified tends to exhibit high perceptual quality, contributing to superior image generation outcomes.

Despite its effectiveness in balancing precision and recall, classifier guidance depends on the gradients of an image classifier, limiting the variability of generated images. This dependency raises the question of whether it is possible to achieve comparable or superior guidance without relying on a classifier.


\textbf{Classifier-free guidance}, introduced by \cite{b7}, addresses this issue by eliminating the need for a dedicated classifier. Instead, it involves training an unconditional denoising diffusion model, parameterized by a score estimator, alongside a conditional denoising diffusion model, parameterized by a conditional score estimator. These are parameterized using a single neural network. For the unconditional model, a null token is used for the class identifier. The models are jointly trained, with the conditioning randomly set to a null token based on a hyperparameter probability p-uncond. This methodology not only simplifies the model architecture but also enhances guidance capabilities without the dependence on an external classifier.

\section{Methodology}\label{sec:methodology}
\begin{figure*}
\centering
\includegraphics[scale=0.5]{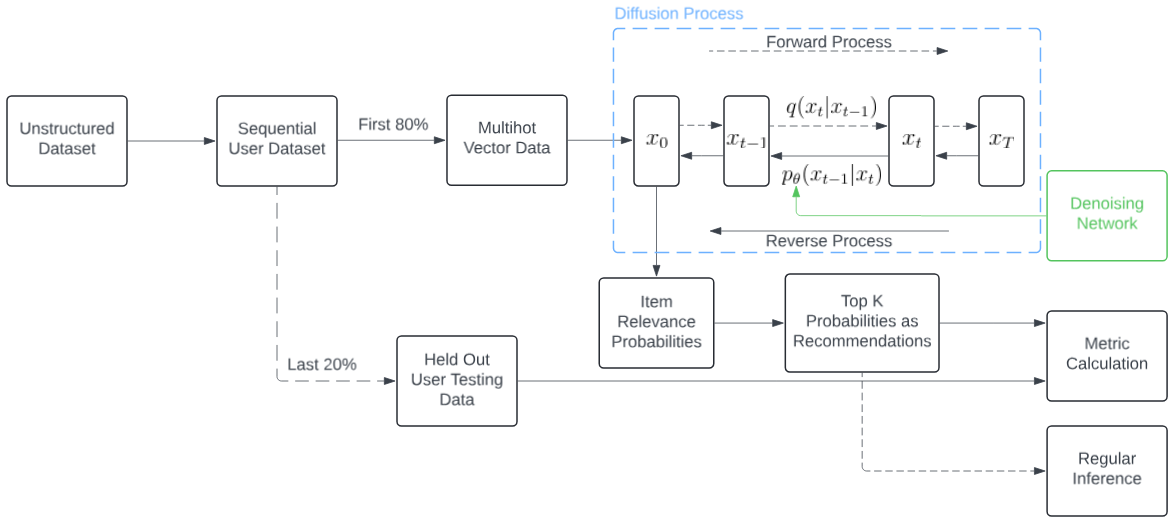}
\caption{Diagram of Our Recommender System}
\label{fig: diagram}
\end{figure*}
The approach adopted in this paper consists of two principle components, detailed in green and blue in Fig.\ref{fig: diagram}. The first component is the diffusion process, encompassing the forward and reverse processes as well as the sampling procedure. Our methodology is inspired by \cite{b15}, which serves as our baseline. In contrast to the original work, which relies solely on partially corrupted original data to guide the denoising model during the reverse process (pseudo-guidance), this study requires an adaptation to implement classifier-free guidance. In this section, we will elucidate the implementation of this external guidance and explore the options available for this type of task. Furthermore, we will discuss the modifications made to the model architecture both independently and as a result of the new form of guidance, and elaborate on the workings of the diffusion process. By thoroughly examining these components, we aim to underscore the necessity and efficacy of incorporating classifier-free guidance into diffusion-based RSs.

The use of partially-noised data for guidance \cite{b15} is inherently incompatible with the classifier-free guidance paradigm. To address this, instead of using partially noised data, we employed pre-noised data to derive guidance. This novel form of guidance, which will be elaborated upon later in this section, enables the seamless integration of classifier-free guidance into the training process. By implementing this revised approach, we aim to enhance the efficacy and flexibility of diffusion-based RSs.

\subsection{Diffusion Process}\label{AA}
The diffusion process consists of a forward process and a reverse process. Given that our modifications primarily impact the reverse process and the denoising model, we will provide an in-depth exploration of the reverse process in the subsequent section. In this section, we will offer a brief overview of how it integrates within the overall diffusion framework. The notations for the forward and reverse processes differ. The forward process is described as follows:

\begin{equation}
q(x_t|x_{t-1}) = \mathcal{N}(x_t, \sqrt{1-\beta_t}x_{t—1},\beta_tI \label{eq1})
\end{equation}


where $x_t$ represents the output of the forward process at step $t$, the output of step $t-1$ ($x_{t-1}$) is the input at time step $t$. $\mathcal{N}$ denotes a normal distribution. In the reverse process, denoted as $p$, we predict the noise to reconstruct $q$, whereas $q$ is a predetermined function for the forward process. For the forward process at timestep $t$, modeled as a normal distribution, we define it by its mean and variance parameters, respectively. $\beta_t$ in both mean and variance refers to a scheduler.

In our work, we employ a linear scheduling approach, which has gained significant traction in recent research. As the term suggests, a linear schedule refers to a sequential method for incrementally adding noise to an image or, in our specific application, to a sequence of interactions. This process can be modeled as a Markov chain, where each step in the sequence is dependent on the previous one to ensure the accurate introduction of noise. Although this iterative approach is characteristic of Markov chains, it could result in inefficiencies during training due to its inherently slow progression. However, this issue is circumvented by adopting the reparameterization technique introduced in the Stable Diffusion framework by \cite{b5}. This method allows us to bypass the need for iterative sampling at each timestep, enabling us to generate the desired output in a single step using the formula:

\begin{equation}
    q(x_t|x_0) = \sqrt{\bar{\alpha}_t}x_0 + \sqrt{1-\bar{\alpha}_t}\epsilon = \mathcal{N}(x_t; \sqrt{\bar{\alpha}_t}x_0, (1-\bar{\alpha}_t)I) \label{eq2}
\end{equation}

With this new formula, we can efficiently calculate the noise to be added at any arbitrary timestep without the need to sequentially traverse all preceding steps. This significantly accelerates the training process.

\subsection{Denoising Model/Reverse Process}
The primary modifications pertain to the denoising model. In this section, we will thoroughly examine these modifications and their implementation. First, we will elucidate the purpose of the denoising model in diffusion models and discuss any differences in its application for RSs. We will then identify the shortcomings of the original denoising model in incorporating classifier-free guidance and explain the steps taken to address these issues.


The denoising model is fundamental to reversing the noise introduced in the forward process. In image diffusion models, a U-Net architecture is commonly employed due to its effectiveness in noise removal tasks. However, given that our dataset is not as extensive as typical image data, we can opt for a simpler architecture, such as a straightforward feedforward network. The primary function of the denoising model is to predict the noise at each timestep, facilitating the reversal process until timestep 0 (fully reversed) is reached. Once the noise is predicted, only a portion of that predicted noise, proportional to the timestep, is removed. Building upon the equation for the reverse process as outlined in the original Stable Diffusion paper \cite{b5} (Eq. \ref{eq3}, \ref{eq4}), our modifications adapt this framework to our specific needs.

\begin{equation}
    p_\theta(x_0:T):=p(x_T)\prod^T_{t=1}p_\theta(x_{t-1}|x_t) \label{eq3}
\end{equation}

\begin{equation}
    p_\theta(x_{t-1}|x_t) = \mathcal{N}(x_{t-1}; \mu_\theta (x_t, t), \sum_\theta (x_t, t)) 
    \label{eq4}
\end{equation}

where $p_\theta (x_0:T)$, known as the diffusion process, is a chain of Gaussian transitions starting at $p(x_T)$ that iterates to $T$ using $p_\theta(x_{t-1}|x_t)$ equation for each singular step. The variance in Eq. \ref{eq4} is time-dependent, but is not trainable. Rather than being set to a constant, it is equal to $\beta_T I$ from the schedule described in the forward process. We refer readers to \cite{b5} for details on mean estimation and other specific aspects. The output for a given timestep $t$ is calculated by:


\begin{equation}
    x_{t-1} = \frac{1}{\sqrt{\alpha t}}(x_t - \frac{\beta_t}{\sqrt{1-\bar{\alpha}}}\epsilon_\theta (x_t,t)) + \sqrt{\beta_t \epsilon}\label{eq5}
\end{equation}

where $\epsilon_\theta (x_t, t)$ is our model’s predicted noise. We simply use this to reverse the diffusion process iteratively using the predicted noise from our denoising model.

The original model architecture employed a straightforward pipeline where the input was passed through an embedding layer and then to an output layer. While this architecture was sufficient for their implementation, we introduced several modifications to enhance performance, particularly with the integration of classifier-free guidance. Firstly, the input to the model was revised to include an unnoised version of the data batch, which serves as guidance for conditioning. Previously, partially noised data acted as pseudo-guidance, but with the new approach, we require explicit external guidance. Therefore, the forward method now takes both the noisy batch of data and the corresponding unnoised batch (hereafter referred to as guidance). Additionally, a parameter, p\_uncond, pertinent to classifier-free guidance, was introduced. This parameter represents the proportion of data where the gradients from the guidance will be zero, effectively enabling non-conditional prediction. In this paper,  $p\_uncond = 0.2$, meaning that 20\% of the guidance data will have zero gradients. The data is then normalized and concatenated with the noised data ($x$). This process trains the denoising model to generate recommendations with and without guidance simultaneously, thereby enhancing the model's capability, as supported by the principles of classifier-free guidance. Although there are various methods to condition the denoising model (e.g., attention mechanisms), we found that the simplest method, concatenation of the data and guidance, was the most effective. The embedding size from the original architecture was maintained, so the concatenated data is transformed to dimensions of (batch size $\times$ embedding size). This transformed data is then run through a hyperbolic tangent activation function and subsequently through another linear layer to project it back to the dimensions corresponding to the number of items in the dataset. This process provides a heatmap of outputs for each item in the dataset. The model remains relatively lightweight, and the rationale behind this design choice will be elaborated upon in Section \ref{sec: config}.

\section{Experimental Setup}\label{sec:setup}

In this section, we will discuss the specifics of the datasets used and how they pertain to our training and testing. We will also discuss the configuration of the experiments we ran when discovering the best method for personalizing recommendations through conditioning and finding the best denoising model architecture. Lastly, the actual results of our chosen architecture and method of guidance in comparison to the original implementation that we based ours upon, as well as the special use cases we have identified that our modified system excels in.

\subsection{Datasets}

We report results for three datasets: the MovieLens 1 Million dataset \cite{b32}, the Yelp dataset \cite{b34}, and several Amazon review datasets \cite{b33}. The inspiration paper specifies two training formats: clean and noisy. While our modifications were tested in both formats, all externally created datasets were limited to the clean format. No significant differences were observed between our modified implementation and the original method for either training format. Datasets statistics are shown in Table \ref{tab: stat}.

For the MovieLens and Yelp datasets, noisy and clean settings are available. However, due to computational constraints, we only employed the MovieLens noisy configuration; the Yelp noisy dataset was excluded. Clean training involves using only ratings of 4 or 5; such ratings are considered ``hits," while ratings of 3 or below are deemed ``misses." Hits are represented as 1 in a vector of length $N$ (number of unique items), whereas misses and unrated items are labeled as 0.

The numeric ratings in the MovieLens 1 Million and Yelp datasets facilitate straightforward classification of hits and misses. In contrast, the Amazon datasets lack numeric ratings, leading us to consider all provided reviews as hits. Although this introduces some noise, it does not significantly affect the differences between our modifications and the original method.

\begin{table}[htbp]
\caption{Dataset Statistics}
\begin{center}
\begin{tabular}{|c|c|c|c|}
\hline
\textbf{Dataset}&\textbf{Users}&\textbf{Items}&\textbf{Total Reviews} \\
\hline
\textbf{ML-1M Clean} & 5949 & 2810 & 403,277\\
\hline
\textbf{ML-1M Noisy} & 5949 & 3494 & 429,993\\
\hline
\textbf{Yelp Clean} & 54,574 & 34,395 & 1,014,486\\
\hline
\textbf{Amazon Kitchen} & 11,566 & 7,722 & 100,464\\
\hline
\textbf{Amazon Beauty} & 3,782 & 2,658 & 38,867\\
\hline
\textbf{Amazon Office} & 1,719 & 901 & 21,466\\
\hline
\textbf{Amazon Toys} & 2,676 & 2,474 & 26,850\\
\hline
\textbf{Amazon VG} & 5,435 & 4,295 & 60,497\\
\hline
\end{tabular}
\label{tab: stat}
\end{center}
\end{table}

The data was split with an 80:20 ratio for training and testing, or 70:20:10 for training, validation, and testing. The last segment of each sequence was reserved for testing. Datasets, primarily containing user ID, item ID, ratings (1-5), review time, and extraneous data, were preprocessed by discarding entries with ratings of 3 or below. Subsequently, the rating column was removed.

For datasets without numeric ratings, all reviews were assumed positive, with allowances for minor noise. Data was sorted by user ID and review time, ensuring sequential interactions per user starting with the oldest. User and item IDs were renumbered from zero to maintain accurate counts post-cleaning.

The preprocessed data was then split into training, validation, and testing sets based on the specified ratios. Post-splitting, the data comprised sequential user interactions, facilitating model training and comparison of recommendations against the most recent user ratings.


The datasets are prepared for training by transforming the pre-runtime format into a more suitable structure. The initial format is inadequate for training since it does not differentiate users effectively, leading to a continuous flow of entries from one user to the next. To address this, we convert the data into a sparse format using \textbf{multi-hot} vectors. Unlike one-hot vectors, which indicate a single hit among items for a specific user, multi-hot vectors represent all hits or interactions with multiple 1’s for each user.

This approach necessitates that the vector length corresponds to the total number of items in the dataset, ensuring accurate item representation. Although the model does not make recommendations based on an exact sequence, it leverages historical data to generate recommendations and uses future data for validation. By preparing the data in this manner, we ensure that the model can effectively differentiate between users and items, thereby facilitating more accurate recommendations. 

\subsection{Experimental Configurations}\label{sec: config}


In this section, we detail the experiments conducted to identify the best-performing recommender system, examining what was successful, what was not, and our reasoning behind these outcomes. Most experiments involved modifications either to the guidance provided to the denoising model or to the denoising model itself. We also conducted minor experiments to explore the potential of certain functionalities.

One key experiment investigated making the recommendation model aware of the actual numeric ratings, rather than using binary signals (1’s and 0’s) to represent hits and misses. In this implementation, negative signals (ratings of 1-3) were introduced alongside positive signals (ratings of 4-5), distinguishing between disliked or unrated items. However, the resulting recommendations were subpar and not comparable to other results, leading us to abandon this approach. The reasons for this performance decline remain unclear.


Through our experiments, we identified a general rule: \textit{a significantly larger parameter count tends to reduce effectiveness}, likely due to overfitting during testing. Given the iterative nature of the denoising process, a large model was unnecessary for the incremental removal of noise from the data. Consequently, configurations with additional layers or increased neurons per layer did not perform well.

We also experimented with attention mechanisms, including self-attention for the input sequence and cross-attentive conditioning from the guidance. Overall, the inclusion of attention did not yield the expected benefits, leading us to opt for a simpler concatenation method to incorporate guidance with input data. This may be attributed to the earlier rule that more parameters do not necessarily enhance performance, and attention mechanisms inherently add parameters.

Different embeddings for the guidance were tested with mixed success. Embeddings proved beneficial primarily with exceptionally large datasets, although they were slightly less effective than using raw data. However, embeddings demonstrated improved efficiency. Therefore, while raw data might be preferable for smaller datasets, embeddings are recommended for scaling the system to handle large datasets. The Yelp dataset, our largest, performed well without embeddings, suggesting that the raw data approach suffices for our current scope.

\subsection{Evaluation} 
For all datasets, we saved the highest-performing model based on the nDCG@10 metric. For evaluation, we computed metrics at K = [1, 5, 10, 20], reflecting the practical importance for recommender systems to generate valuable recommendations within this range. We initially considered higher values but concluded that average users are unlikely to search through 50 or 100 recommendations.
All configurations were trained for varying durations depending on the dataset size but extended well beyond the point of reaching the best-performing results on the testing datasets.

\section{Experimental Results}\label{sec:results}

\begin{table}[]
\caption{Testing Results. Org denotes original results. Values are in percentage.}
\label{tab:ml-yelp}
\begin{tabular}{|c|cc|cc|cc|}
\hline
\multirow{2}{*}{Dataset} & \multicolumn{2}{c|}{ML-1M Clean} & \multicolumn{2}{c|}{ML-1M Noisy} & \multicolumn{2}{c|}{Yelp Clean} \\ \cline{2-7} 
 & \multicolumn{1}{c|}{Original} & Ours & \multicolumn{1}{c|}{Original} & Ours & \multicolumn{1}{c|}{Original} & Ours \\ \hline
\multirow{4}{*}{\begin{tabular}[c]{@{}c@{}}Precision\\ @\\ {[}1, 5, 10, 20{]}\end{tabular}} & \multicolumn{1}{c|}{8.64} & \textbf{9.25} & \multicolumn{1}{c|}{2.99} & \textbf{4.63} & \multicolumn{1}{c|}{2.8} & \textbf{2.61} \\
 & \multicolumn{1}{c|}{7.92} & \textbf{8.61} & \multicolumn{1}{c|}{3.37} & \textbf{3.6} & \multicolumn{1}{c|}{2.22} & 2.08 \\
 & \multicolumn{1}{c|}{7.16} & \textbf{7.83} & \multicolumn{1}{c|}{3.46} & \textbf{3.57} & \multicolumn{1}{c|}{1.9} & 1.87 \\
 & \multicolumn{1}{c|}{6.54} & \textbf{6.82} & \multicolumn{1}{c|}{3.45} & \textbf{3.55} & \multicolumn{1}{c|}{1.61} & \textbf{1.61} \\ \hline
\multirow{4}{*}{\begin{tabular}[c]{@{}c@{}}Recall\\ @\\ {[}1, 5, 10, 20{]}\end{tabular}} & \multicolumn{1}{c|}{0.69} & \textbf{0.92} & \multicolumn{1}{c|}{0.29} & \textbf{0.61} & \multicolumn{1}{c|}{0.48} & \textbf{0.48} \\
 & \multicolumn{1}{c|}{3.18} & \textbf{3.96} & \multicolumn{1}{c|}{1.51} & \textbf{2.22} & \multicolumn{1}{c|}{1.85} & 1.77 \\
 & \multicolumn{1}{c|}{5.49} & \textbf{6.6} & \multicolumn{1}{c|}{3.04} & \textbf{3.7} & \multicolumn{1}{c|}{3.12} & 3.11 \\
 & \multicolumn{1}{c|}{9.72} & \textbf{11.66} & \multicolumn{1}{c|}{5.55} & \textbf{6.57} & \multicolumn{1}{c|}{5.14} & \textbf{5.19} \\ \hline
\multirow{4}{*}{\begin{tabular}[c]{@{}c@{}}nDCG\\ @\\ {[}1, 5, 10, 20{]}\end{tabular}} & \multicolumn{1}{c|}{8.64} & \textbf{9.25} & \multicolumn{1}{c|}{2.99} & \textbf{4.63} & \multicolumn{1}{c|}{2.8} & 2.61 \\
 & \multicolumn{1}{c|}{8.33} & \textbf{9} & \multicolumn{1}{c|}{3.49} & \textbf{4.18} & \multicolumn{1}{c|}{2.67} & 2.53 \\
 & \multicolumn{1}{c|}{8.43} & \textbf{9.32} & \multicolumn{1}{c|}{3.97} & \textbf{4.58} & \multicolumn{1}{c|}{2.94} & 2.88 \\
 & \multicolumn{1}{c|}{9.44} & \textbf{10.56} & \multicolumn{1}{c|}{4.83} & \textbf{5.57} & \multicolumn{1}{c|}{3.56} & 3.53 \\ \hline
\multirow{4}{*}{\begin{tabular}[c]{@{}c@{}}MRR\\ @\\ {[}1, 5, 10, 20{]}\end{tabular}} & \multicolumn{1}{c|}{8.64} & \textbf{9.25} & \multicolumn{1}{c|}{2.99} & \textbf{4.63} & \multicolumn{1}{c|}{2.8} & 2.61 \\
 & \multicolumn{1}{c|}{15.41} & \textbf{16.39} & \multicolumn{1}{c|}{6.78} & \textbf{8.27} & \multicolumn{1}{c|}{5.29} & 4.94 \\
 & \multicolumn{1}{c|}{17.06} & \textbf{18.27} & \multicolumn{1}{c|}{8.34} & \textbf{9.73} & \multicolumn{1}{c|}{6.04} & 5.77 \\
 & \multicolumn{1}{c|}{18.15} & \textbf{19.33} & \multicolumn{1}{c|}{9.38} & \textbf{10.9} & \multicolumn{1}{c|}{6.62} & 6.35 \\ \hline
\end{tabular}
\end{table}

\begin{table*}[]
\centering
\caption{Results for Amazon dataset. Values are in percentage.}
\label{tab:amazon}
\begin{tabular}{|c|cc|cc|cc|cc|cc|}
\hline
\multirow{2}{*}{Dataset} & \multicolumn{2}{c|}{Amazon Kitchen} & \multicolumn{2}{c|}{Amazon Beauty} & \multicolumn{2}{c|}{Amazon Toys} & \multicolumn{2}{c|}{Amazon Office} & \multicolumn{2}{c|}{Amazon VG} \\ \cline{2-11} 
 & \multicolumn{1}{c|}{Original} & Ours & \multicolumn{1}{c|}{Original} & Ours & \multicolumn{1}{c|}{Original} & Ours & \multicolumn{1}{c|}{Original} & Ours & \multicolumn{1}{c|}{Original} & Ours \\ \hline
\multirow{4}{*}{\begin{tabular}[c]{@{}c@{}}Precision \\      @\\      {[}1, 5, 10, 20{]}\end{tabular}} & \multicolumn{1}{c|}{0.35} & \textbf{0.52} & \multicolumn{1}{c|}{0.93} & \textbf{1.06} & \multicolumn{1}{c|}{1.12} & 0.75 & \multicolumn{1}{c|}{0.29} & \textbf{0.58} & \multicolumn{1}{c|}{1.47} & 1.29 \\ \cline{2-11} 
 & \multicolumn{1}{c|}{0.21} & \textbf{0.24} & \multicolumn{1}{c|}{0.87} & \textbf{0.90} & \multicolumn{1}{c|}{0.45} & 0.30 & \multicolumn{1}{c|}{0.35} & \textbf{0.52} & \multicolumn{1}{c|}{0.90} & \textbf{1.09} \\ \cline{2-11} 
 & \multicolumn{1}{c|}{0.14} & \textbf{0.16} & \multicolumn{1}{c|}{0.83} & 0.81 & \multicolumn{1}{c|}{0.32} & 0.26 & \multicolumn{1}{c|}{0.26} & \textbf{0.61} & \multicolumn{1}{c|}{0.90} & 0.84 \\ \cline{2-11} 
 & \multicolumn{1}{c|}{0.10} & \textbf{0.10} & \multicolumn{1}{c|}{0.62} & \textbf{0.67} & \multicolumn{1}{c|}{0.20} & \textbf{0.23} & \multicolumn{1}{c|}{0.17} & \textbf{0.44} & \multicolumn{1}{c|}{0.70} & \textbf{0.71} \\ \hline
\multirow{4}{*}{\begin{tabular}[c]{@{}c@{}}Recall \\      @\\      {[}1, 5, 10, 20{]}\end{tabular}} & \multicolumn{1}{c|}{0.13} & \textbf{0.39} & \multicolumn{1}{c|}{0.38} & \textbf{0.42} & \multicolumn{1}{c|}{0.42} & \textbf{0.47} & \multicolumn{1}{c|}{0.04} & \textbf{0.10} & \multicolumn{1}{c|}{0.82} & 0.72 \\ \cline{2-11} 
 & \multicolumn{1}{c|}{0.42} & \textbf{0.56} & \multicolumn{1}{c|}{1.64} & \textbf{1.71} & \multicolumn{1}{c|}{0.97} & 0.64 & \multicolumn{1}{c|}{0.74} & 0.65 & \multicolumn{1}{c|}{2.32} & \textbf{2.63} \\ \cline{2-11} 
 & \multicolumn{1}{c|}{0.62} & \textbf{0.68} & \multicolumn{1}{c|}{3.44} & 3.18 & \multicolumn{1}{c|}{1.15} & \textbf{1.36} & \multicolumn{1}{c|}{1.08} & \textbf{1.77} & \multicolumn{1}{c|}{4.63} & 4.12 \\ \cline{2-11} 
 & \multicolumn{1}{c|}{0.89} & 0.79 & \multicolumn{1}{c|}{5.19} & \textbf{5.28} & \multicolumn{1}{c|}{1.37} & \textbf{2.20} & \multicolumn{1}{c|}{1.34} & \textbf{2.57} & \multicolumn{1}{c|}{6.80} & \textbf{6.99} \\ \hline
\multirow{4}{*}{\begin{tabular}[c]{@{}c@{}}nDCG \\      @\\      {[}1, 5, 10, 20{]}\end{tabular}} & \multicolumn{1}{c|}{0.35} & \textbf{0.52} & \multicolumn{1}{c|}{0.93} & \textbf{1.06} & \multicolumn{1}{c|}{1.12} & 0.75 & \multicolumn{1}{c|}{0.29} & \textbf{0.58} & \multicolumn{1}{c|}{1.47} & 1.29 \\ \cline{2-11} 
 & \multicolumn{1}{c|}{0.35} & \textbf{0.55} & \multicolumn{1}{c|}{1.37} & \textbf{1.42} & \multicolumn{1}{c|}{0.94} & 0.65 & \multicolumn{1}{c|}{0.61} & \textbf{0.61} & \multicolumn{1}{c|}{1.94} & \textbf{2.03} \\ \cline{2-11} 
 & \multicolumn{1}{c|}{0.41} & \textbf{0.59} & \multicolumn{1}{c|}{2.05} & 1.92 & \multicolumn{1}{c|}{0.98} & 0.89 & \multicolumn{1}{c|}{0.72} & \textbf{1.06} & \multicolumn{1}{c|}{2.75} & 2.54 \\ \cline{2-11} 
 & \multicolumn{1}{c|}{0.50} & \textbf{0.64} & \multicolumn{1}{c|}{2.58} & 2.57 & \multicolumn{1}{c|}{1.04} & \textbf{1.15} & \multicolumn{1}{c|}{0.82} & \textbf{1.35} & \multicolumn{1}{c|}{3.40} & \textbf{3.49} \\ \hline
\multirow{4}{*}{\begin{tabular}[c]{@{}c@{}}MRR \\      @\\      {[}1, 5, 10, 20{]}\end{tabular}} & \multicolumn{1}{c|}{0.35} & \textbf{0.52} & \multicolumn{1}{c|}{0.93} & \textbf{1.06} & \multicolumn{1}{c|}{1.12} & 0.75 & \multicolumn{1}{c|}{0.29} & \textbf{0.58} & \multicolumn{1}{c|}{1.47} & 1.29 \\ \cline{2-11} 
 & \multicolumn{1}{c|}{0.53} & \textbf{0.70} & \multicolumn{1}{c|}{1.94} & 1.76 & \multicolumn{1}{c|}{1.54} & 0.87 & \multicolumn{1}{c|}{0.95} & \textbf{1.09} & \multicolumn{1}{c|}{2.52} & \textbf{2.61} \\ \cline{2-11} 
 & \multicolumn{1}{c|}{0.57} & \textbf{0.74} & \multicolumn{1}{c|}{2.22} & 2.14 & \multicolumn{1}{c|}{1.59} & 1.02 & \multicolumn{1}{c|}{1.05} & \textbf{1.42} & \multicolumn{1}{c|}{3.07} & 2.94 \\ \cline{2-11} 
 & \multicolumn{1}{c|}{0.61} & \textbf{0.77} & \multicolumn{1}{c|}{2.36} & 2.32 & \multicolumn{1}{c|}{1.61} & 1.14 & \multicolumn{1}{c|}{1.10} & \textbf{1.54} & \multicolumn{1}{c|}{3.37} & 3.27 \\ \hline
\end{tabular}
\end{table*}

Some of the results from our experiments yield intriguing insights into specific scenarios where our model excels, particularly when compared to other methods and the original implementation. The hyperparameter configuration remained the same as in the original work, except for the addition of the classifier-free guidance hyperparameter $p\_uncond$ set at 0.2. The results are shown in Tables \ref{tab:ml-yelp} an \ref{tab:amazon}, in which any bold value represents a score that either surpasses or ties with the other implementations.

Our system generally performs on par with, or better than, the original method across most scenarios. When our method is outperformed, the margin is typically small. Conversely, when our method outperforms the original, the margin is substantially larger, outperforming the original by more than 100\% in some situations. This trend holds for nearly all scenarios, with only occasional outliers. One such outlier is the Yelp dataset with clean data. Our model fails to beat the original approach on this dataset systematically, we suspect that this is due to the method of applying guidance. For the majority of our datasets, a complex embedding to encapsulate the information of the guidance matrix is not needed, however, the Yelp dataset may be one such scenario as it is noticeably larger. The experimentation into an efficient embedding model for guidance fell out of the scope of this paper. Overall, our approach equals or surpasses the original implementation in 85 out of 128 categories across all datasets, indicating a clear improvement.

We also believe that our recommendation system, enhanced by classifier-free guidance, could excel in few-shot and potentially zero-shot scenarios. This hypothesis is supported by the marked improvements observed in the Amazon Office dataset, which has a limited item count and thus offers less personalization. Generally, we found our model to work better in scenarios where either fewer items were offered for personalization or there were fewer reviews per user. Classifier-free guidance involves training without guidance for 20\% of the samples, creating a scenario analogous to zero-shot learning, which might contribute to its effectiveness in this area and would explain its significant performance in comparison to the original method when less guiding data is available. Although we did not explore this aspect in-depth, our preliminary results suggest potential benefits in few-shot and zero-shot scenarios.

\section{Conclusion and Future Work}\label{sec:conclusion}

This paper presents classifier-free guidance as a form of personalization and incorporates denoising in diffusion-based recommender frameworks. By restructuring the guidance mechanism by removing partially noised pseudo-guidance and incorporating true guidance through conditioning on pre-noise data, we demonstrated improvements in performance. The effectiveness of our method was evaluated using precision, recall, nDCG, and mean reciprocal rank @ K on held-out user data, revealing clear benefits over previous methods. Additionally, our approach showed potentially enhanced capabilities in few-shot and zero-shot recommendation scenarios.

Our major contributions include: (1) the implementation and evaluation of classifier-free guidance in diffusion-based recommendation tasks; (2) an in-depth discussion on the parameterization and architecture of the denoising model utilized in these tasks; and (3) insights into the applicability of diffusion models beyond the visual domain. Our findings indicate that classifier-free guidance improves performance in recommendation tasks similarly to its success in visual tasks. Additionally, we provide a framework that addresses challenges within diffusion-based recommendation systems while highlighting the newfound capabilities in few-shot and zero-shot scenarios.

The potential for future work in this area is substantial. One promising direction involves improving the system by considering actual numerical ratings. While our initial attempt at this was inconclusive, a more sophisticated approach, potentially involving triplet loss to account for negative reviews, could yield better results. Cross-attentive conditioning, another feature that did not meet our performance expectations, might benefit from refined architectural innovations for improved guidance.

The use of transformers in the diffusion process is another avenue worth exploring. Recent advancements in image generation have shown that replacing the U-Net backbone with a Vision Transformer (ViT) can enhance denoising capabilities, and similar benefits might be realized in recommendation systems. Additionally, the few-shot and zero-shot capabilities of diffusion recommender systems trained with classifier-free guidance warrant further investigation. This aspect is particularly relevant for e-commerce, where new users with no interaction history present ongoing challenges.

In conclusion, our study lays the groundwork for advancing diffusion-based recommendation systems. By addressing the identified shortcomings and exploring the outlined directions for future research, we can further enhance these systems to meet the evolving needs of recommender applications.

\end{document}